\documentclass[pre,aps,amsmath,amssymb,showkeys,12pt]{article}
\usepackage{graphicx,color,amssymb,amsmath,epsfig}
  \makeatletter
	\renewcommand\@biblabel[1]{#1.}
	\makeatother
	
	\makeatletter
	\def\@cite#1#2{$(#1\if@tempswa , #2\fi)$}
	\makeatother

\nonstopmode 

\long\def\symbolfootnote[#1]#2{\begingroup%
\def\thefootnote{\fnsymbol{footnote}}\footnote[#1]{#2}\endgroup}

\begin{document}
\addvspace{6cm}
\begin{center}{\Huge{The Vaccinee's Dilemma and Influenza Epidemics}}
\\
\large{Raffaele Vardavas,$^\ast$ Romulus Breban,\symbolfootnote[1]{R.V.~and R.B.~have contributed equally to the research described in this paper.} ~Sally Blower\symbolfootnote[2]{To whom correspondence should be addressed. E-mail: sblower@mednet.ucla.edu}}
\\
\large{Semel Institute for Neuroscience and Human Behavior, University of California, Los Angeles, California 90095-1555, USA.}
\end{center}
%\date{}
\newpage

{\bf
Inspired by Minority Games, we constructed a novel individual-level game of adaptive decision-making based on the dilemma of deciding whether to participate in voluntary influenza vaccination programs. The proportion of the population vaccinated (i.e., the vaccination coverage) determines epidemic severity. Above a critical vaccination coverage, epidemics are prevented; hence individuals find it unnecessary to vaccinate. The adaptive dynamics of the decisions directly affect influenza epidemiology and, conversely, influenza epidemiology strongly influences decision-making. This feedback mechanism creates a unique self-organized state where epidemics are prevented.  This state is attracting, but unstable; thus epidemics are rarely prevented. This result implies that vaccination will have to be mandatory if the public health objective is to prevent influenza epidemics. We investigated how collective behavior changes when public health programs are implemented. Surprisingly, programs requiring advance payment for several years of vaccination prevents severe epidemics, even with voluntary vaccination. Prevention is determined by the individuals' adaptability, memory, and number of pre-paid vaccinations. Notably, vaccinating families exacerbates and increases the frequency of severe epidemics. }
\\
game theory $|$ agent-based modeling $|$ complex social dynamics $|$ vaccines 
 
\pagebreak

\section*{Introduction} 
Since the influenza vaccine is effective for one year, individuals must decide every year whether to participate in voluntary vaccination programs \cite{Szucs}. The proportion of the population vaccinated, $p$ (i.e., the vaccination coverage) determines epidemic severity \cite{AM}. Above a critical vaccination coverage $p_c$, epidemics are prevented; hence individuals find it unnecessary to vaccinate. Individuals that vaccinate protect themselves from infection. If they do not vaccinate they may still avoid infection if sufficient numbers of their peers vaccinate (i.e., through {\it herd immunity}). This poses a yearly dilemma (i.e., the {\it vaccinee's dilemma}) for the individual of whether vaccination is necessary. An individual's strategy for making yearly decisions may be an adaptive process of trial and error based on their decision making experience (i.e., using {\it inductive reasoning}). We model the adaptive dynamics of vaccination decisions in a population of non-communicating individuals acting in their own self-interest trying to avoid infection preferably without vaccinating. We track individual-level decisions and population-level variables; specifically, yearly vaccination coverage and prevalence $f$ (i.e., the proportion of the population infected during an influenza season). We address the following question:
{\it Can an ensemble of individuals self-organize to prevent influenza epidemics by yearly voluntary vaccination?}	

Mean-field influenza transmission models (based on ordinary differential equations) have explored the epidemiological effects of treatment \cite{Longini}, drug resistance \cite{Stilianakis}, multiple strains \cite{Gog}, and virulence \cite{Mills} at the population-level. Spatial data have been used to forecast influenza epidemics \cite{Valleron1}, and investigate synchronization between epidemics \cite{Valleron2}. In the past decade, Minority Games \cite{ZhangBook} have attracted much attention in statistical mechanics and finance. Minority Games model the behavior of adaptive individuals within a collective that compete for the benefit of being in a minority.  Inspired by Minority Games, we constructed an individual-level game of adaptive vaccination decision making. Our  individual-level game includes the Susceptible-Infected-Recovered model (see Supplementary Methods Appendix A) to determined whether the critical vaccination coverage necessary to prevent influenza epidemics is likely to be reached. Previously, games have been used to price vaccines \cite{Geo} and predict voluntary vaccination coverage for pathogens that provide permanent immunity (e.g., smallpox or measles) \cite{Earn1}. These games are based on deductive reasoning. For pathogens that provide permanent immunity, deductive reasoning can be used because individuals need to get vaccinated only once.  In the case of influenza, individuals do not acquire permanent immunity and need to make vaccination decisions every year. Thus, it may be assumed that individuals make decisions based on their past experiences (i.e., use inductive reasoning) rather than based only on the current epidemiology (i.e., use deductive reasoning). In contrast to previous work, our model is an inductive reasoning game.  The central concept of our model is that the adaptive dynamics of the vaccination decisions directly affect influenza epidemiology and, conversely, influenza epidemiology strongly influences decision-making. 

\section*{Results}
We find that influenza epidemics cannot be prevented in most seasons if individuals make decisions based upon inductive reasoning and voluntary vaccination (Fig.~1A). When epidemics occur, some individuals get infected and hence increase their vaccination probability for the next season; thus, $p$ approaches $p_c$ (Fig.~1A). Eventually, the coverage slightly exceeds $p_c$ due to the stochastic nature of the individual-level vaccination decisions and an epidemic is prevented. Next season, many individuals decide that there is no need for vaccination; thus $p$ abruptly decreases and a severe epidemic ensues. (This effect resembles the occurrence of crashes in financial markets \cite{Sornette} due to sudden loss of public confidence in the stock exchange.) Coverage then repeats similar dynamics (Fig.~1A). Yearly prevalence peaks every time coverage drops substantially below $p_c$, rapidly decreasing in successive years as coverage increases (Fig.~1A). If initial coverage is larger than $p_c$ then $p$ will drop below $p_c$ within a few years (results not shown). Coverage then follows similar dynamics to that illustrated in Fig.~1A. The feedback mechanism between the influenza epidemiology and the adaptive dynamics of the decision making creates a unique self-organized state where epidemics are prevented.  This state is attracting, but unstable; thus epidemics are rarely prevented. This result implies that vaccination will have to be mandatory if the public health objective is to prevent influenza epidemics. In the Supplementary Methods, we show the fixed point analysis and derive the expected periodicity of the coverage dynamics.    

The dynamics of each individual's vaccination probability is more complex than coverage dynamics (Fig.~1B). Figure 1C shows two distributions for the individual's probabilities of vaccination $w^{(i)}$. The first distribution (black) is obtained when epidemics are prevented. Individuals segregate into two groups as found for other inductive reasoning games \cite{Johnson}. Individuals in one group are very likely to vaccinate whilst individuals in the other group are very likely not to vaccinate, few are undecided. The second distribution (blue) is obtained in the successive seasons when severe epidemics occur. In these seasons, the distribution of the vaccination probabilities remains segregated. However, individuals who were very likely to vaccinate decrease their vaccination probability (Fig.~1C) causing severe epidemics. As $p$ increases back to $p_c$, the second distribution (blue) slowly tends towards the first distribution (black). Eventually, $p$ again exceeds $p_c$ and the distribution undergoes a similar cyclic dynamics.

We then investigated how collective behavior changes if certain public health programs are implemented, but vaccination remains voluntary. Specifically, we focused on whether either of two programs (based upon incentives) would prevent influenza epidemics. For the first program we assumed that if the head of the family pays for his/her vaccination, then their family would get vaccinated for free. The head of the family uses inductive reasoning to make decisions and modifies his/her probability of vaccinating next season depending upon the number of family members that were infected. Surprisingly, this program exacerbated epidemics and also increased the frequency of severe epidemics, even with voluntary vaccination. Coverage dynamics appear similar to the original dynamics with no program; Fig.~2A. However, $p$ dropped more frequently below $p_c$ and thus more severe epidemics occurred; the time average of the coverage $\left< p \right>$ is lower than in the absence of the program. This program exacerbated epidemics because it reduced the number of independent decision makers; see Supplementary Methods for details.

For the second public health program we assumed that free vaccination would be offered for $y$ number of successive years if the individual paid for vaccination in the first year, and that individuals who participate in the program take the offer of free vaccination for all $y$ successive years. Individuals do not make decisions during the years of free vaccination, but they still evaluate the necessity of vaccination. They then use their evaluations to decide whether to pay for a subsequent vaccination and receive a further $y$ free vaccinations. This program (even though it is based upon voluntary vaccination) could greatly ameliorate epidemics. Amelioration depend upon the: memory parameter $s$, adaptability parameter $\epsilon$, and number of years of free vaccination $y$. Figure 2B shows coverage dynamics for $y=3$ years (red -- the epidemic is ameliorated) and $y=15$ years (green -- the epidemic is exacerbated); $y=0$ (black) is shown for comparison. For $y=3$ years $\left <p \right >$ increases; epidemics are much less severe but more frequent than for $y=0$ years. Furthermore, the average prevalence $\left < f \right >$ is substantially lower than for $y=0$ years. For $y=15$ years epidemics occur with greater severity; moreover average prevalence for $y=15$ years is greater than for $y=0$ years. Epidemics were prevented most seasons because average coverage $\left <p \right >$ exceeded the critical coverage $p_c$. Figure 3 shows the parameter regions for memory $s$ and adaptability $\epsilon$ that are important for increasing $\left <p \right >$ and decreasing $\left < f \right >$. When $y=0$, epidemics were ameliorated if the population was composed of individuals who readily adapted their decisions and/or who did not quickly forget their past vaccination experience (Figs.~3Ai and 3Aii). The situation is more complex when $y>0$; there are disconnected areas in the $(\epsilon,s)$ parameter space where the program can exacerbate epidemics; Fig.~3.

\section*{Discussion} 
In the United States (US), demand for influenza vaccines is generally met and no major shortages occur. In recent years, coverage (based upon voluntary vaccination) has steadily increased \cite{CDC3,CDC}. One of the national health objectives of the US is to increase the coverage \cite{CDC3,CDC} that is currently below the Healthy People 2010 objectives \cite{HP2010}. We found complex epidemic dynamics can occur simply due to individuals making voluntary vaccination decisions. We deliberately did not include variation in strain virulence because we wanted to assess the impact of voluntary vaccination on pathogen dynamics. Notably, we found large influenza epidemics could occur without the introduction of virulent pandemic strains. Furthermore, the critical coverage level is unlikely to be reached if vaccination is voluntary. We also found that certain public health programs could be detrimental. Previous studies have shown that vaccinating above the critical coverage level would always be worthwhile. In contrast, we paradoxically found that a coverage level higher than the critical level could be problematic. This perverse result occurs because inductively reasoning individuals evaluate vaccination as unnecessary if the critical coverage level was exceeded in the previous season. The severity of the subsequent influenza epidemic depends upon the degree to which the critical coverage was exceeded.
  
We used a novel individual-level inductive reasoning game model to obtain insight into the impact of individual-level vaccination decisions on influenza epidemiology. We found that critical vaccination coverage levels will rarely be reached; hence influenza epidemics will only occasionally be prevented. This result implies that vaccination will have to be mandatory if the public health objective is to prevent influenza epidemics. The two central assumptions of our model are that individuals do not communicate their vaccination decisions and act in their own self-interest. If other assumptions are made then other outcomes are possible. If a herding mechanism \cite{Zimmermann} occurs where individuals share their perception of risk of getting infected, then the public may vaccinate even if when it does not appear worthwhile for them; this behavior may prevent epidemics. However, even if individuals act in the interest of their own families epidemics would not be prevented if vaccination is voluntary. In contrast, public health programs, based on pre-payment, could greatly ameliorate influenza epidemics.

\section*{Methods}

In the past decade, inductive reasoning systems have been modeled using Minority Games \cite{ZhangBook,Arthur}. A Minority Game models how non-communicating selfish individuals reach collective behavior with respect to a common dilemma under adaptation of each one's expectations \cite{ZhangBook}. Minority Games have been used to model financial markets \cite{Challet2}. Here we present the first vaccination decision model based on Minority Game methodology. We model a population of $N$ individuals acting in their own self-interest that do not communicate their vaccination decisions with each other.  Every year, these individuals independently decide whether or not to vaccinate against influenza. Individuals use their past experiences to help them make their vaccination decisions for the current season.  The probability that individual $i$ chooses to vaccinate in season $n$ is $w^{(i)}_n$. Individual $i$ is assigned a variable $V^{(i)}_n$ the value of which depends upon whether or not: the individual choses to vaccinate, they became infected, and an epidemic occurred (Fig.~4A). $V^{(i)}_n$ increases each time the individual perceives that there was, or would have been, a benefit to vaccination because (a) the individual vaccinated and there was an epidemic, or (b) the individual did not vaccinate and then became infected (Fig.~4A).  

We model the effect of memory by using a parameter $s$ to discount the previous seasons' vaccination outcome with respect to the outcome of the present season ($0< s< 1$). Specifically, $V^{(i)}_{n+1}= s V^{(i)}_n+1$ if individual $i$ believes he did, or would have, benefited from vaccination in season $n$. Otherwise, if individual $i$ believes that vaccination was unnecessary in season $n$ (regardless of whether he got vaccinated or not), $V^{(i)}_{n+1}= s V^{(i)}_n$. We normalize $V^{(i)}_{n+1}$ by $(s^{n+1}-1)/(s-1)$ because this factor is the maximum possible value for $V^{(i)}_{n+1}$ if individual $i$ would have benefited from vaccination in all $n$ influenza seasons. The domain for $V^{(i)}$ is $0\leq V^{(i)} < 1/(1-s)$. The probability that an individual chooses to vaccinate in the next influenza season is given by $w^{(i)}_{n+1}= (1-\epsilon) w^{(i)}_n+\epsilon V^{(i)}_{n+1}/[(s^{n+1}-1)/(s-1)]$, where $\epsilon$ describes the adaptability with respect to experience with vaccination ($0< \epsilon < 1$).
			
In our model, the probability of an unvaccinated individual acquiring influenza $q(p)$ decreases linearly as the coverage $p$ increases; Fig.~4B.  When $p \geq p_c$ the probability of an unvaccinated individual getting infected is defined to be zero. This linear function is a good approximation of the relationship found for the Susceptible-Infected-Recovered model; see Supplementary Methods Appendix A. Our model describes a large population of individuals. Thus, we account only for epidemics and we do not consider outbreaks. Outbreaks become decreasingly important as the population size $N$ increases. Although we do not include the option of treatment, the effects of treatment can be implicitly accounted in our model by decreasing $p_c$.

At the end of the season, each individual evaluates their vaccination decisions based upon whether vaccination was necessary to avoid infection. Consequently, they modify their probability to vaccinate next season to $w^{(i)}_{n+1}$. Figure 4A shows a diagram of the evaluation tree. Individuals start their first season with no prior experience in decision making as whether to vaccinate or not. The initial condition assigns a random vaccination probability for the first season to every individual. Specifically, $V^{(i)}_0=0$ and $w^{(i)}_0=U[0,1)$ for all individuals where $U[0,1)$ is a uniformly random variable between $0$ and $1$. Our initial conditions are chosen to reflect the fact that the initial public awareness of the benefits of influenza vaccination would not be high enough to prevent an epidemic, while at the individual-level, the likelihood of vaccination may vary considerably. 
  
  For the case of the first public health program, based upon voluntary vaccination, we considered a population of $N$ individuals that are grouped into families with $C$ members. The head of the family updates his/her $V^{(j)}_{n}$ value
  (where $j$ labels the family, i.e., $j=1...N/C$) in the following way:  (a) {$V^{(j)}_{n+1}= s V^{(j)}_n+C$ if the head of  the family had decided to vaccinate his/her family and there was an epidemic that season;} (b) {$V^{(j)}_{n+1}= s V^{(j)}_n$ if there was no epidemic that season, regardless of whether or not the family was vaccinated against influenza;} (c) {$V^{(j)}_{n+1}= s V^{(j)}_n+k$ if $k$ members of the family were infected in a season where  the head of the family decided not to vaccinate his/her family.}  We normalized the value of  $V^{(j)}_{n+1}$ by a factor of $C(s^{n+1}-1)/(s-1)$ that represents the  maximum possible value of $V_{n+1}^{(j)}$. Therefore, the vaccination probabilities
  are updated as follows $w^{(j)}_{n+1}= (1-\epsilon) w^{(j)}_n+\epsilon  V^{(j)}_{n+1}/[C (s^{n+1}-1)/(s-1)]$. 
\newline

The authors gratefully acknowledge financial support from NIH/NIAID (RO1 AI041935). We also thank Nelson Freimer, Tara Martin, Justin Okano, and David Wilson for discussions during the course of this research.

{\bf Supplementary Information accompanies the paper.}

\pagebreak

\newpage
\section*{}

\pagebreak

\begin{figure}
\centering
%\resizebox{14cm}{!}{\includegraphics{fig1.eps}}
\caption{Vaccination dynamics using a memory parameter $s=0.7$, an adaptability parameter $\epsilon=1$, a critical coverage $p_c=0.6$ (dashed line) and a probability $q(0)=0.8$ of getting infected when $p=0$. \textbf{A}: Dynamics of yearly coverage $p$ for a population of $N= 10^5$ individuals (black), and the corresponding dynamics of the prevalence $f$ (red). The dynamics of $p$ is approximately cyclic: as $p$ approaches $p_c$ from below, eventually fluctuates above $p_c$, and then abruptly drops. \textbf{B}: $w^{(i)}_n$ versus time for two individuals in the population.  In contrast to the simple dynamics of the coverage, individuals go through complex decision behavior. \textbf{C}: Normalized distributions $\rho(w^{(i)})$ of $w^{(i)}_n$ for a population of $N=10^7$ individuals. The distribution when the coverage fluctuates above $p_c$ is shown in black and the successive distribution when the coverage abruptly drops below $p_c$ is shown in blue.  Individuals tend to strongly segregate into two groups. The individuals in the first group are highly likely not to vaccinate next season. In the black distribution, the individuals in the second group are highly likely to vaccinate (i.e., $w=1$). In the blue distribution the individuals in the second group are less likely to vaccinate than previously (i.e., $w=s$, given that no epidemic occurred in the previous season).} 
\label{fig:1}
\end{figure}

\begin{figure}
\centering
%\resizebox{9cm}{!}{\includegraphics{fig2.eps}}
\caption{Coverage dynamics for different public heath programs (that are based on voluntary vaccination) in a population of $N=10^5$ individuals using a memory parameter $s=0.7$, an adaptability parameter $\epsilon=1$, a critical coverage $p_c=0.6$, and a probability $q(0)=0.8$ of getting infected when $p=0$. \textbf{A}: The head of the family makes the decision as to whether or not their family vaccinates. The coverage dynamics when the family size is eight ($C=8$) is shown in blue; the coverage dynamics when individuals make vaccination decisions independently is shown in black for comparison. Similar results were obtained for family sizes of two and four.  \textbf{B}: Individuals that pay for one vaccination are rewarded with $y=3$ (red) and $y=15$  (green) extra years of vaccination; the coverage dynamics when individuals pay for every year of vaccination is shown in black for comparison.} 
\label{fig:2}
\end{figure}

\begin{figure}
\centering
%\resizebox{53mm}{!}{\includegraphics{fig3A.eps}}\resizebox{45mm}{!}{\includegraphics{fig3B.eps}}\resizebox{45mm}{!}{\includegraphics{fig3C.eps}}
\caption{Structure of the average prevalence $\left <f\right>$ (panels {\bf i}) and the average coverage $\left <p\right>$ (panels {\bf ii}) in the parameter space $(\epsilon,s)$. \textbf{A}: no public health program -- i.e., $y=0$ (gray scale). \textbf{B}: Individuals that pay for one vaccination are rewarded with $y=3$ extra years of vaccination (red scale). \textbf{C}: Individuals that pay for one vaccination are rewarded with $y=15$ extra years of vaccination (green scale).} 
\label{fig:3}
\end{figure}

\begin{figure}
\centering
%\resizebox{14cm}{!}{\includegraphics{fig4.eps}}
\caption{\textbf{A}: Diagram illustrating the evaluation tree. An individual that decides to vaccinate (branch (a)) will judge  their decision depending on whether there was an influenza epidemic that season.  If $p\geq p_c$ (branch (a1)), they will conclude that  getting vaccinated that season was not necessary to prevent infection. Otherwise, if $p<p_c$ (branch (a2)), they will conclude that  their decision was beneficial for avoiding infection that season. An individual that decides not to vaccinate that season (branch (b)) will judge  their decision based on whether or not they were infected. If they did get infected (branch (b1)) they will conclude that  their decision of not vaccinating was detrimental and that vaccination was necessary for avoiding infection. Instead, if by chance they avoided infection (branch (b2)), they will conclude that vaccination was not necessary. \textbf{B}: The probability of getting infected with influenza $q(p)$ versus the vaccination coverage $p$.} 
\label{fig:4}
\end{figure}

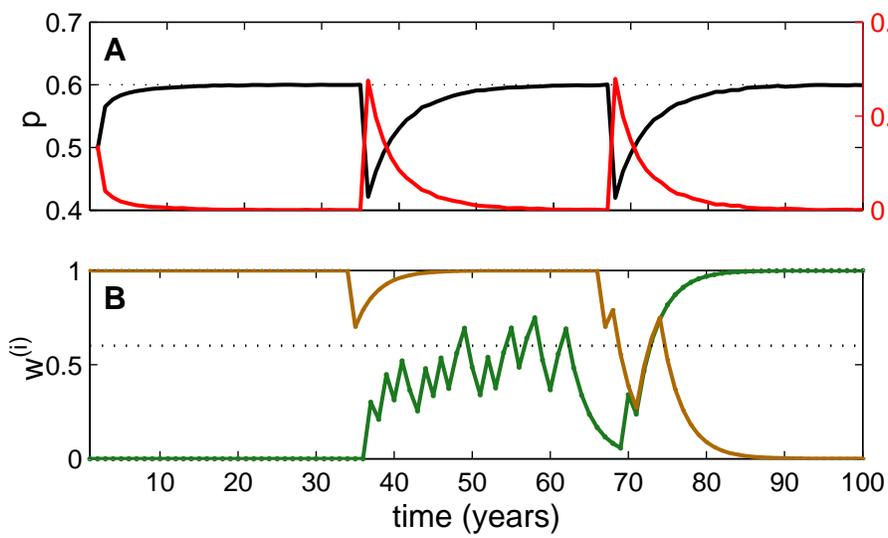
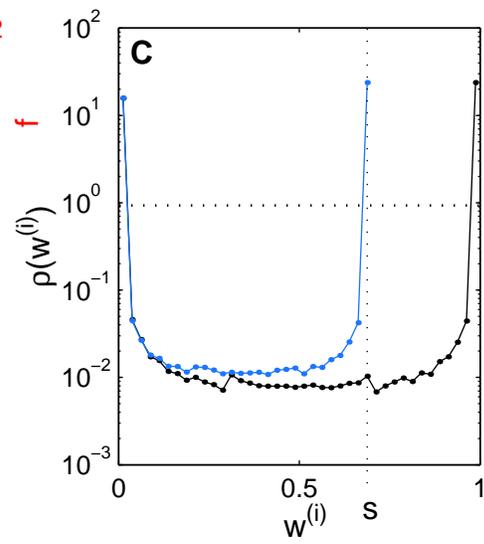
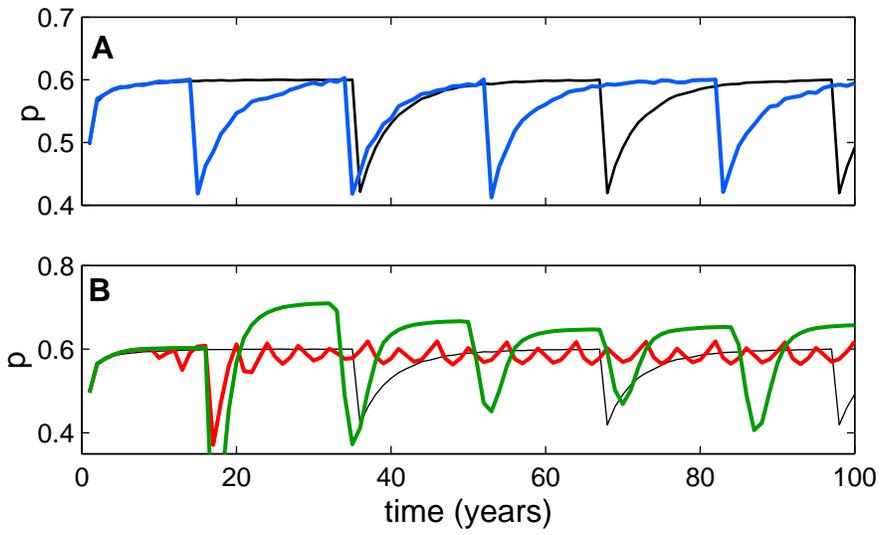
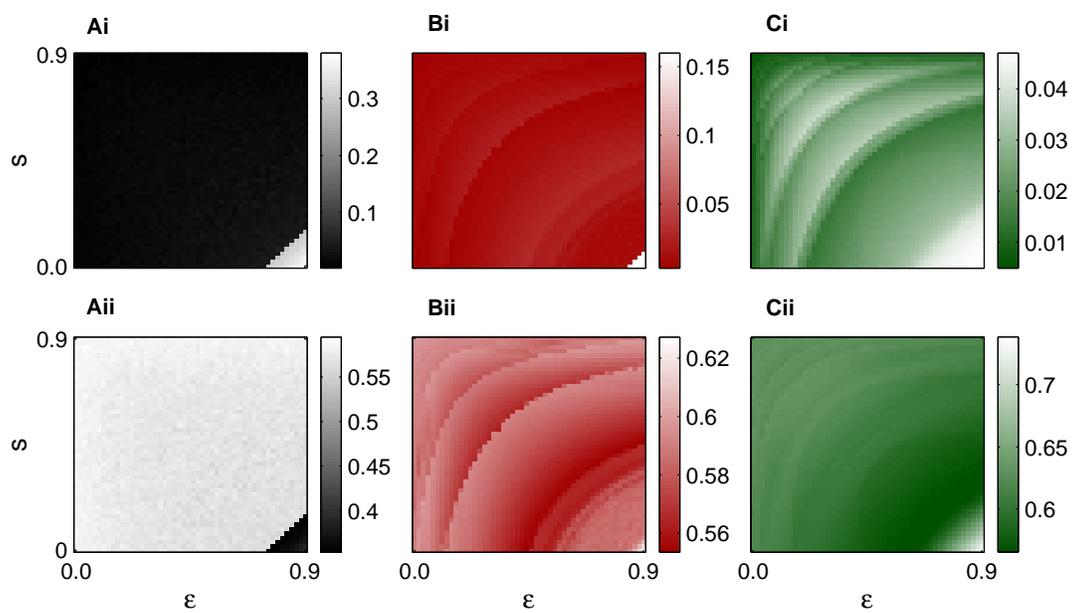
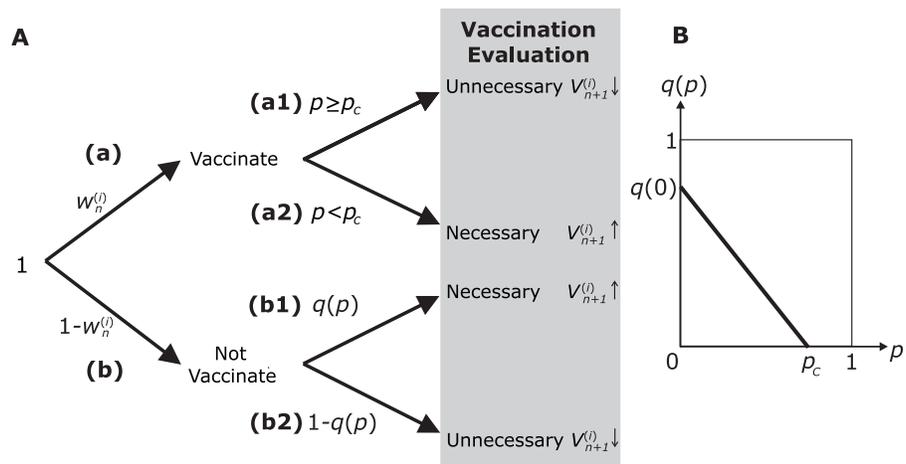
\end{document}